\documentclass[pdflatex,sn-mathphys-num]{sn-jnl}

\usepackage{graphicx}%
\usepackage{multirow}%
\usepackage{amsmath,amssymb,amsfonts}%
\usepackage{amsthm}%
\usepackage{mathrsfs}%
\usepackage[title]{appendix}%
\usepackage{xcolor}%
\usepackage{textcomp}%
\usepackage{manyfoot}%
\usepackage{booktabs}%
\usepackage{algorithm}%
\usepackage{algorithmicx}%
\usepackage{algpseudocode}%
\usepackage{listings}%
\usepackage{csquotes}
\usepackage{tabularx,ragged2e}
\newcolumntype{L}{>{\RaggedRight}X}
\newcolumntype{C}{>{\centering}X}

\begin{document}

\title[Article Title]{Retrospective of the ARPA-E BETHE-GAMOW-Era Fusion Programs and Project Cohorts}


\author*[1,5]{\fnm{S. C.} \sur{Hsu}}\email{scott@lowercarbon.com}
\author[2,5]{\fnm{M. C.} \sur{Handley}}
\author[3,5]{\fnm{S. E.} \sur{Wurzel}}
\author[4,5]{\fnm{P. B.} \sur{McGrath}}

\affil[1]{\orgname{Lowercarbon Capital}, \orgaddress{\city{Los Angeles}, \state{CA} \postcode{90034}, \country{USA}}}
\affil[2]{\orgname{Strong Atomics}, \orgaddress{\city{San Francisco}, \state{CA}, \postcode{94110},  \country{USA}}}
\affil[3]{\orgname{Fusion Energy Base}, \orgaddress{\city{New York}, \state{NY} \postcode{10003},\country{USA}}}
\affil[4]{\orgname{Schmidt Family Foundation}, \orgaddress{\city{Palo Alto}, \state{CA} \postcode{94301}, \country{USA}}}
\affil[5]{This is an account of work performed while the authors, who are listed with their present affiliations, were formerly with ARPA-E\@. Additional significant contributors, who are presently federal employees or contractors, are listed in the acknowledgments. This paper represents the views of the authors and does not necessarily represent the views of the U.S. Government}


\abstract{This paper provides a retrospective of the BETHE (Breakthroughs Enabling THermonuclear-fusion Energy) and GAMOW (Galvanizing Advances in Market-aligned fusion for an Overabundance of Watts) fusion programs of the Advanced Research Projects Agency-Energy (ARPA-E), as well as fusion project cohorts (associated with OPEN~2018, OPEN~2021, and Exploratory Topics) initiated during the same time period (2018--2022). BETHE (announced in 2019) aimed to increase the number of higher-maturity, lower-cost fusion approaches. GAMOW (announced in 2020) aimed to expand and translate research-and-development efforts in materials, fuel-cycle, and enabling technologies needed for commercial fusion energy. Both programs had a vision of enabling timely commercial fusion energy while laying the foundation for greater public-private collaborations to accelerate fusion-energy development. Finally, this paper describes ARPA-E's fusion Technology-to-Market (T2M) activities during this era, which included supporting ARPA-E fusion performers' commercialization pathways, improving fusion costing models, exploring cost targets for potential early markets for fusion energy, engaging with the broader fusion ecosystem (especially investors and nongovernmental organizations), and highlighting the importance of social license for timely fusion commercialization.}



\maketitle

\section*{Introduction}\label{sec1}

A previous paper~\cite{nehl19} describes the origins of fusion at ARPA-E (Advanced Research Projects Agency-Energy) of
the U.S. Department of Energy (DOE) and provides a retrospective of ARPA-E's first fusion program ALPHA~\cite{alpha} (Accelerating Low-cost Plasma Heating and Assembly). 
This paper provides a retrospective of the next era of ARPA-E fusion programs and projects initiated between 2018--2022, centering on the BETHE~\cite{bethe} (Breakthroughs Enabling Thermonuclear-fusion Energy, pronounced ``beta'') and GAMOW~\cite{gamow} (Galvanizing Advances in Market-aligned fusion for an Overabundance of Watts) programs, as well as several fusion project ``cohorts'' associated with OPEN 2018, OPEN 2021, and Exploratory Topics, during the same era. GAMOW was a joint program with Fusion Energy Sciences (FES) of the DOE Office of Science (SC). The total obligated federal funding for the programs and project cohorts covered
in this paper is approximately \$108M as of May~2025.

ARPA-E programs are driven by its Program Directors. The seeds of the BETHE-GAMOW-era programs and project cohorts were encapsulated in the first author's
testimony to the Energy Subcommittee of the Committee on Science, Space, and Technology
of the U.S. House of Representatives in April 2016~\cite{hsu16testimony}. Salient 
points of the testimony, which underpinned the BETHE-GAMOW-era programs, included: 
\begin{itemize}
    \item There is a wide range of scientifically credible, though lower scientific-maturity, fusion concepts that could potentially lower the cost and shorten the timeline
    to commercial fusion energy;
    \item A range of such concepts were supported in the past by the DOE but ceased being supported c.~2010;
    \item Congress and DOE should re-assess innovative fusion concept development and implement a program with appropriate program and project metrics to support timely development toward economically competitive fusion energy;
    \item Support the development and use of tools (e.g., computational codes, diagnostic capabilities, domestic and international facilities) and engineering solutions (e.g., plasma-facing and tritium-breeding systems) needed by many fusion concepts;
    \item Enable near-term public-private partnerships with private fusion companies.
\end{itemize}

Context for the development of the BETHE and GAMOW programs at \mbox{ARPA-E} 
c.~2019 was anchored in
the identified global need for $\sim$500~exajoules per year of sustainable, carbon-neutral primary energy by mid-century, with just a handful of energy technologies having the potential to contribute to this need in the foreseeable future: renewables plus long-duration storage; nuclear fission; fossil fuels with carbon capture, utilization, sequestration (CCUS); and enhanced geothermal. We argued that adding fusion energy to this mix
could ease all paths to reaching net-zero carbon emissions, and that fusion may represent the most sustainable energy source for the very long term, especially when considering the attributes of energy security and abundance.

Fusion-specific context 
for the development of the BETHE and GAMOW programs was built around
the following narrative. Fusion was on the cusp of demonstrating the major milestone of
energy gain (see \cite{wurzel22} for various definitions), but world fusion research programs
were focused on scientific advances and were not on a trajectory to  
impact energy markets and carbon-emissions reductions by mid-century. Meanwhile, the
scientific and technological (S\&T) maturity of the materials and enabling technologies (e.g., tritium-breeding and fuel-cycle technologies) needed for a fusion
plant was far behind that of the plasma fusion source. Privately funded
efforts (about US\$1.5 billion of cumulative private equity investments in
fusion companies at that time) were taking higher-scientific-risk
approaches to commercial fusion energy development, leveraging
the latest S\&T capabilities and innovations. While the significant private investments
demonstrated market interest and emphasized commercialization relevance, we felt that it would be extremely challenging
if not impossible for any company to realize a grid-ready demonstration
with private funding alone. Expanding upon the changing narratives around fusion initiated in part by the ALPHA program, we argued that 
ARPA-E should help fusion break out from being a purely big-science enterprise, consistent
with the second of two recommendations from the 2019 U.S. National Academies report \textit{A Strategic Plan for U.S. Burning Plasma Research}~\cite{NASEM19}: “the U.S. should start a national program of accompanying research and technology leading to the construction of a compact pilot plant that produces electricity from fusion at the lowest possible capital cost.” 

\begin{figure}[h]
\centering
\includegraphics[width=0.9\textwidth]{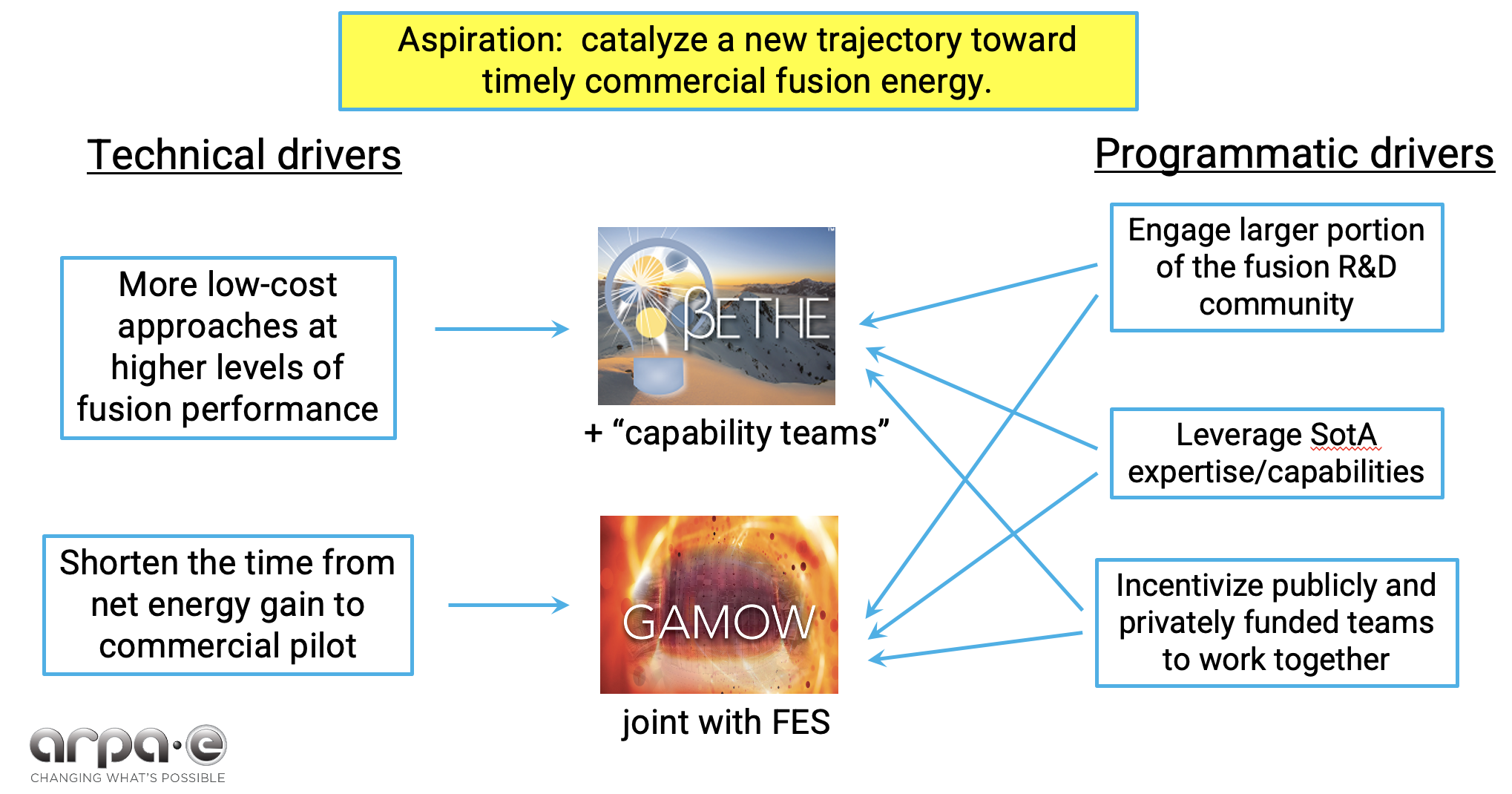}
\caption{Graphic shown at the 2022 ARPA-E Fusion Annual Meeting (April 26--27, 2022 in
San Francisco, CA) summarizing the technical and programmatic drivers of BETHE and GAMOW. The latter was released as a joint program with Fusion Energy Sciences (FES) of the DOE Office of Science.}\label{fig:context}
\end{figure}

Figure~\ref{fig:context} shows the high-level BETHE-GAMOW-era
technical and programmatic objectives.
Table~\ref{table1} provides a summary of the fusion programs and cohorts during the BETHE-GAMOW era. Further details about each program/cohort are provided below in their respective sections. 
It is worth acknowledging that nearly all the BETHE-GAMOW-era 
project teams initiated and/or conducted the bulk of their
work amidst the global COVID pandemic, contending with limited ability
to perform experimental work, as well as significant supply-chain disruptions and procurement delays. We
express our sincere gratitude to all the dedicated project performers
for the significant progress made despite these challenges.

The remainder of the paper is organized under the following sections:
``BETHE,'' ``GAMOW,'' ``Other Fusion Project Cohorts,'' ``BETHE-GAMOW-Era T2M Activities,'' ``Conclusions,'' and ``Appendix~A: Select Project T2M Notes.''

\begin{table}[h]
\caption{Summary of BETHE-GAMOW-era fusion programs and project cohorts.}\label{table1}%
\begin{tabularx}{\textwidth}{>{\hsize=.26\hsize}X>{\hsize=.1\hsize\centering\arraybackslash}X>{\hsize=.38\hsize}X>{\hsize=.1\hsize\centering\arraybackslash}X>{\hsize=.1\hsize\centering\arraybackslash}X}
\toprule
Program & Year announced & Theme & \# of projects & Funding\footnotemark[1]\\
\midrule
OPEN 2018~\cite{FOA_OPEN2018} & 2017 & Concept development & 3 & \$11.48M\\
Fusion Diagnostics~\cite{FOA_TINA} & 2019 & Transportable diagnostics & 8 & \$8.01M\\
BETHE~\cite{FOA_BETHE} & 2019 & Concept/component development & 18 & \$48.40M\footnotemark[2]\\
GAMOW~\cite{FOA_GAMOW} & 2020 & Materials \& enabling technologies & 14 & \$29.64M\footnotemark[3]\\
OPEN 2021~\cite{FOA_OPEN2021} & 2021 & Multiple & 4 & \$10.66M\\
\botrule
\end{tabularx}
\footnotetext[1]{Total obligated federal funding as of March~2025.}
\footnotetext[2]{Includes \$4.52M from SC-FES for distinct project scope added during award negotiations.}
\footnotetext[3]{Jointly funded by SC-FES.}
\end{table}

\section*{BETHE}\label{sec:BETHE}

\subsection*{BETHE Program Objective}

The BETHE program \cite{bethe} was informed in part by an ARPA-E program-development workshop
entitled ``Enabling Timely and Commercially Viable Fusion Energy,''
held near San Francisco on August 13--14, 2019.
The BETHE Funding Opportunity Announcement (FOA)~\cite{FOA_BETHE}, released on Nov.~7, 2019, stated:

\begin{displayquote}
{\em The program objective is to help enable an eventual grid-ready fusion demonstration with OCC\footnote{Overnight capital cost, assuming 10th-of-a-kind for fusion-specific systems and $n$th-of-a-kind for balance-of-plant systems} $<$US\$2B and $<$\$5/W by (1)~developing and advancing the performance of multiple lower-cost (see Section I.E of the FOA for cost metrics) fusion concepts, and (2) reducing the capital cost of more-mature, higher-cost concepts through development and application of new, innovative component technologies. ARPA-E believes that the existence of a larger number of lower-cost fusion concepts above threshold values of verified performance (see Section I.E of the FOA for milestones and metrics), which is lacking today, could attract further private investments and build the foundation for a sustainable public-private partnership to develop fusion energy. In this partnership scenario, riskier, early-stage development is supported primarily through federal funding, while performance and engineering scale-up is supported primarily through private investments.}
\end{displayquote}

\begin{displayquote}
{\em ARPA-E recognizes that a number of significant scientific and technical challenges remain to be solved before a grid-ready fusion demonstration is possible. It must be emphasized that ARPA-E does not expect any project in this program to achieve net-energy breakeven, let alone a grid-ready demonstration, before the end of the program. Specifically, this program aims to enable at least two or three lower-cost concepts to reach $T_e,T_i \ge 1$~keV for the first time (for those concepts), where $T_e$ and $T_i$ are the electron and ion temperatures, respectively. Similarly, the program aims to enable at least one concept to achieve fusion triple product $nT\tau_E \ge 10^{18}$~keV$\cdot$s/m$^3$ for the first time (for that concept). These performance metrics represent a significant reduction in scientific risk for fusion concepts and, thus, a more attractive reward-to-risk ratio for private investors. In addition, this program aims to develop and demonstrate prototype component technologies that could significantly reduce the cost of two or three higher-cost, more-mature fusion concepts.}
\end{displayquote}

\begin{figure}[h]
\centering
\includegraphics[width=0.9\textwidth]{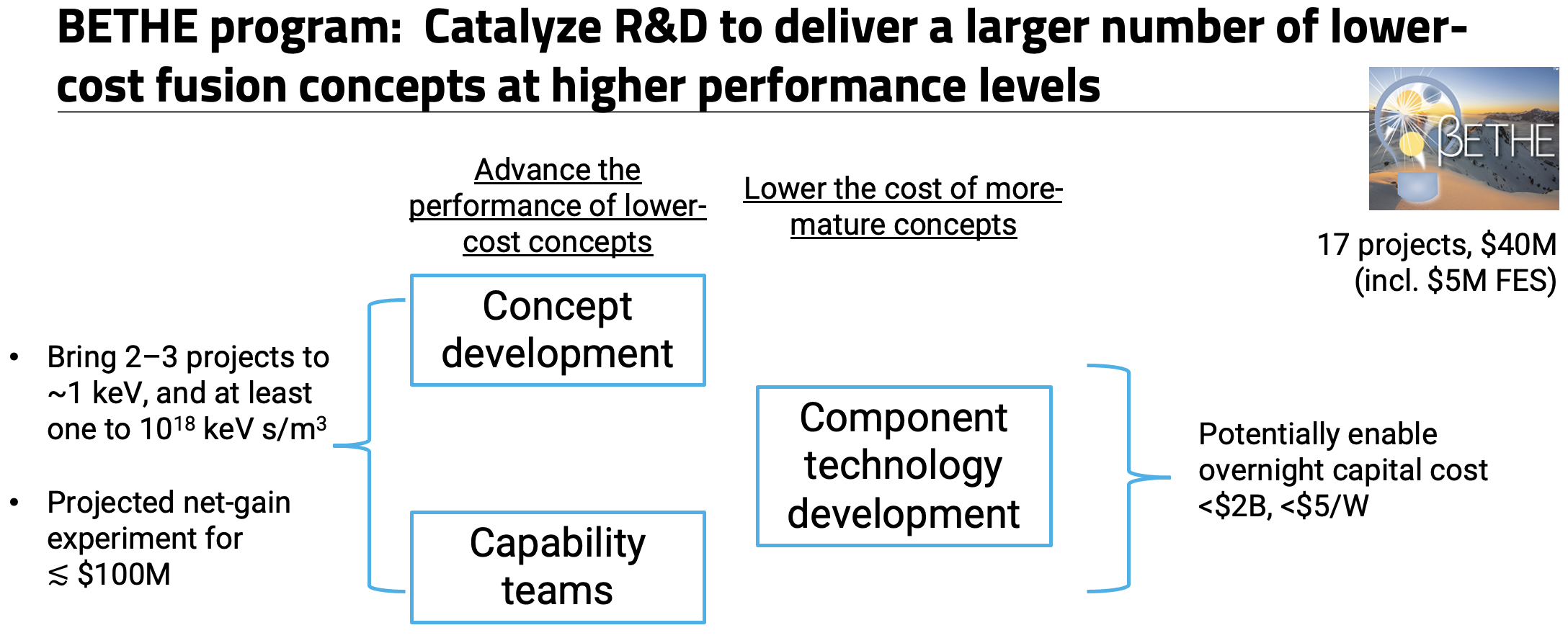}
\caption{Graphic shown at the 2022 ARPA-E Fusion Annual Meeting (April 26--27, 2022 in
San Francisco, CA) summarizing the objectives and technical categories of the BETHE program.}\label{fig:BETHE}
\end{figure}

Figure~\ref{fig:BETHE} summarizes BETHE's objective and technical categories, which are described in detail below.

\subsection*{BETHE Technical Categories}

To achieve the program objective, BETHE applicants were requested to select one of three technical research categories: (A)~concept development, (B)~component technology development, or (C)~capability teams, as described below.

For Category~A (Concept Development), applicants were invited to mature the performance of lower-cost fusion concepts. In particular,
applicants were required to provide a cost estimate of a net-energy-gain\footnote{The FOA clarified this to be ``wall-plug'' gain (see~\cite{wurzel22}), which
is an aggressive definition seeking to elevate concepts with lower
development needs with respect to plasma formation,
heating, and/or sustainment technologies.}
experiment, i.e., short pulse for magnetic-confinement-fusion (MCF) concepts or single-shot for inertial-confinement-fusion (ICF) or magneto-inertial-fusion (MIF) concepts. A costing spreadsheet, based on~\cite{sheffield16},
was provided to help applicants estimate the costs of components/subsystems each expected to cost more than 5\% of the total OCC\@. The more that the spreadsheet result exceeded \$100M, the greater a detailed
justification was needed to explain how the high-level metric of OCC $<$US\$2B and $<$\$5/W for a
grid-ready demonstration could still be met. Any concept that required more than \$10M of funding (not including non-federal cost share) to achieve a successive technical milestone (see next subsection for defined milestones) was not eligible for this category. The FOA stated that ARPA-E anticipated supporting a range of fusion concepts and maturity levels, with commensurate budgets, from the full range of fusion parameter space~\cite{lindemuth09} (i.e., from low-density magnetically confined up to high-density, inertially confined concepts). This indicated ARPA-E's continued expansion (first seen in OPEN 2018) in supporting fusion concepts beyond the pulsed, intermediate-density concepts of ALPHA\@. Category-A applications were limited to \$10M/project, not including cost share.

For Category~B (Component Technology Development), applicants were invited to develop component technologies, up to prototype
demonstration, that could significantly reduce the capital
cost of a specific, more-mature, higher-cost fusion concept, such as but not limited to the
tokamak, stellarator, reversed-field pinch (RFP), or laser-driven inertial confinement fusion (ICF)\@.
Applicants were required to provide a quantitative argument (e.g., relative to a relevant past system
study) that a grid-ready demonstration power plant enabled by the new technology could
potentially satisfy the metric of OCC $<$US\$2B and $<$\$5/W for a grid-ready demonstration. Category-B applications were limited to \$4M/project, not including cost share.

For Category~C (Capability Teams), applicants were invited to improve, adapt, and apply existing capabilities (including
theory/modeling, creative application of artificial intelligence or machine learning, engineering support or fabrication, or 
diagnostics) to support multiple teams in the first two categories and to lower overall program costs. Capability teams were expected to be expert teams, including multiple institutions if appropriate,
with demonstrated experience in their proposed capability. They were not to be advocates of a particular fusion concept.
This category was envisioned to build a strong foundation for fusion public-private partnerships. Category-C applications were limited to \$2.5M/project, not including cost share.

\subsection*{BETHE Technical Performance Targets}

\subsubsection*{Category A: Concept Development}

Applicants were requested to advance the performance of their proposed fusion concept based on defined technical milestones,
as shown in Table~2 (which is an abbreviated version of Table~1 of the BETHE FOA~\cite{FOA_BETHE}). Applicants were required to enter the program at the highest milestone previously
achieved for the concept (not necessarily achieved by the applicant), and aim to exit at any higher milestone, as long as the
funding limit (not including non-federal cost share) was not exceeded. Concepts based on hypothetical low-energy nuclear reactions (LENR), aka ``cold fusion,'' were not eligible (ARPA-E later released an FOA for an Exploratory Topic on LENR~\cite{FOA_LENR}).
Any applicants proposing concepts involving appreciable non-thermal plasmas were required to explain how energy gain may be achieved despite the large energy inputs required to sustain a non-thermal energy distribution.

\begin{table}[h]
\caption{Entry (already achieved) and aspirational exit (at project end) milestones that BETHE Category-A applicants were required to select, as well as the application funding ceiling (not including non-federal cost share) for achieving each successive milestone.}\label{table2}%
\begin{tabularx}{\textwidth}{@{} cLc @{}}
\toprule
Milestone & Description & Funding ceiling \\	
\midrule
1 & Theoretical analysis showing that net energy gain is possible & \$1M \\
2 & Physics-based simulations indicating plausibility of commercially relevant energy gains & \$5M \\
3 & Demonstration of plasma configuration suitable for scale-up in $nT\tau$\footnotemark[1]
& \$5M \\
4 & Plasma with $T_i$, $T_e \ge 100$~eV & \$10M \\
5 & Plasma with $T_i$, $T_e \ge 1$~keV and/or $nT\tau \ge 10^{18}$~keV$\cdot$s/m$^3$ & \$10M \\
6 & Plasma $nT\tau\ge 10^{19}$~keV$\cdot$s/m$^3$ & \$10M \\
7 & Plasma $nT\tau\ge 10^{20}$~keV$\cdot$s/m$^3$ & N/A \\
\botrule
\end{tabularx}
\footnotetext[1]{Fusion triple product of the fuel density $n$, temperature $T$, and confinement time $\tau$. See~\cite{wurzel22} for details. As the FOA \cite{FOA_BETHE} indicated, this metric is meaningful only for thermal fusion plasmas.}
\end{table}

The FOA provided further discussion with
respect to how ARPA-E would judge that a milestone has been or will be achieved. 
For milestones 1--2, this would require sound theoretical analysis 
progressing to physics-based numerical simulations. Milestones 3 and above would require convincing diagnostic data,
typically including density, temperature, and magnetic-field measurements (if applicable) with
rudimentary information on spatial profiles and temporal evolution. The upper milestones would require inference of
confinement time ($\tau$ or $\tau_E$ depending on concept, see~\cite{wurzel22}) and measurement of neutron yield. The FOA specified that neutron yield 
could not be the only measurement, and that evidence/plans should be provided to differentiate between thermal and
beam-generated neutrons. Finally, Category-A applicants were strongly
encouraged to coordinate with Capability Teams (see corresponding subsection below) starting at the proposal stage, and to clarify 
which topics would be excluded in any potential collaboration from
the perspective of intellectual-property (IP) protection. 

\subsubsection*{Category B: Component Technology Development}

Applicants for Category B were requested to provide a quantitative argument (based on analysis or in a cited reference) on how their proposed, innovative component
technology could lower the cost of a grid-ready demonstration based on a specific higher-performing
fusion concept such as (but not limited to) the tokamak, stellarator, RFP, or laser-driven ICF, to
meet the metrics of OCC $<$US\$2B and $<$\$5/W for a grid-ready fusion demonstration. The
analysis could have been based on updating one or more past cost studies
relevant to the proposed concept.

Anticipating widely varying research topics for Category B, the FOA did
not provide a universal set of performance targets
against which progress would be measured. Applicants were are asked to self-identify critical-path
milestones directly tied to risk reduction in developing the innovative component
technology, and to provide quantitative metrics in the application.
Applicants were also requested to clearly state the desired outcome at the end of the project with
quantitative metrics and the next research-and-development (R\&D) step that the outcome would enable.

\subsubsection*{Category C: Capability Teams}

Anticipating applicants with many different
topics/capabilities, it was impossible for the FOA to state universal performance targets for Category C\@.
Applicants were requested to clearly state what capability they intended to provide
to the program, what they planned to do for Category-A concept teams, and what specific
tools, codes, or resources they would be using. Applicants were
required to provide a high-level plan that
included any required preparatory or development tasks as well as a notional
schedule for working with multiple potential Category-A project teams.
ARPA-E expected that Capability Teams and
Concept Development Teams would reach mutually acceptable agreements on data rights and IP
protection, and strongly encouraged publication of scientific results that did not reveal protected
information.

\subsection*{BETHE Awardees and Outcomes}

There was an overwhelming response to the BETHE FOA~\cite{FOA_BETHE}, leading eventually
to 17 competitively selected awardees~\cite{bethe}, as summarized in Table~\ref{table3}, plus
a T2M costing project discussed below in the T2M section. It is beyond the scope of
this paper to include a detailed summary of each project (brief
descriptions of BETHE projects are available at \cite{bethe}) and its outcomes, although we
provide citations to published papers for each project in 
column~2 of Table~3. Below, we provide high-level comments on 
the program outcomes, as well as that of each technical category relative to the stated program objectives.

\begin{table}[!ht]
\caption{Summary of BETHE awardees with citations to published S\&T results.}\label{table3}%
\begin{tabularx}{\textwidth}{>{\hsize=.21\hsize}X>{\hsize=.52\hsize}X>{\hsize=.19\hsize}X>{\hsize=.08\hsize\centering\arraybackslash}X}
\toprule
Lead institution & Project title &  & Funding \\
\midrule
Category A & & Concept & \\
\midrule
Univ.\ Wisconsin-Madison
& An HTS Axisymmetric Magnetic Mirror on a Faster Path to Lower Cost Fusion Energy \cite{egedal22,ruegsegger23,radovinsky23,endrizzi23,ialovega23,forest24,kristofek25,tran25,frank25}
& Axisymmetric mirror & \$10.30M \\
Zap Energy & Sheared Flow Stabilized Z-Pinch Performance Improvement \cite{meier21,banasek23,levitt23,levitt24,datta24,goyon24,crews24} & Stabilized Z-Pinch & \$1.00M \\ 
Univ.\ Maryland, Baltimore County & Centrifugal Mirror Fusion Experiment \cite{schwartz24} & Centrifugal mirror & \$5.18M \\
NK Labs & Conditions for High-Yield Muon-Catalyzed Fusion & Muon-catalyzed fusion & \$2.03M \\
Univ.\ Washington & Demonstration of Low-Density, High-Performance Operation of Sustained Spheromaks and Favorable Scaling toward Compact, Low-Cost Fusion Power Plants \cite{morgan22,hossack22} & Sustained spheromak & \$1.50M \\
LANL & Target Formation and Integrated Experiments for Plasma-Jet Driven Magneto-Inertial Fusion (PJMIF) \cite{chu23prl,lajoie23a,lajoie23b,vyas23,chu23pop,lajoie24} & PJMIF & \$5.79M \\ 
\midrule
Category B & & Component & \\
\midrule
Commonwealth Fusion Systems (CFS) & Pulsed High Temperature Superconducting (HTS) Central Solenoid (CS) for Revolutionizing Tokamaks \cite{lee22,huang23a,huang23b,huang23c,sanabria24,watterson25} & HTS-CS (tokamak) & \$2.91M\footnotemark[1] \\
PPPL & Stellarator Simplification using Permanent Magnets & Magnet array for 3D fields (stellarator) & \$3.34M\footnotemark[1] \\
Univ.\ Rochester \& NRL (co-leads) & Advanced Inertial Fusion Energy (IFE) Target Designs and Driver Development \cite{schmitt23a,schmitt23b,bates23,petrova23,lawrence23,myers24,wolford25} & Diode \& excimer lasers (IFE) & \$6.14M\footnotemark[1]\\
Type One Energy & Demonstration of High Temperature Superconducting Non-Planar Stellarator Magnet with Advanced Manufactured Assemblies \cite{riva23} & Bent HTS magnets (stellarator) & \$1.18M \\
\midrule
Category C & & Capability & \\
\midrule
Virginia Tech. & Capability in Theory, Modeling, and Validation for a Range of Innovative Fusion Concepts using High-Fidelity Moment-Kinetic Models \cite{cagas21,mathews21,cagas23,skolar23,francisquez23,bradshaw24,bradshaw25} & Modeling & \$2.40M \\
Sapientai & Data-Enabled Fusion Technology \cite{oliver24} & ML/AI & \$1.65M \\
Univ.\ Rochester & A Simulation Resource Team for Innovative Fusion Concepts \cite{davies21,walsh21,wen22,diaw22,garcia22,calvo23,davies23a,davies23b,hansen24,lavell24a,sefkow24,garcia24,lavell24b,lawrence24,michta24,lavell25} & Modeling & \$2.25M\\
MIT & Radio Frequency Scenario Modeling for Breakthrough Fusion Concepts & Modeling & \$1.25M \\
ORNL & Magnetic Field Vector Measurements Using Doppler-Free Saturation Spectroscopy & Diagnostics & \$600k\\
LANL & Electromagnetic and Particle Diagnostics for Transformative Fusion-Energy Concepts \cite{lajoie24,klemmer24} & Diagnostics & \$425k \\
Woodruff Sci.\footnotemark[2] & Fusion Costing Study and Capability & Cost modeling & \$450k \\
\botrule
\end{tabularx}
\footnotetext[1]{Includes SC-FES funding of \$1.54M for CFS, \$342k for PPPL, and \$2.64M for Rochester and NRL.}
\footnotetext[2]{Funded through BETHE T2M with PPPL.}
\end{table}

BETHE's overarching technical and programmatic objectives are
shown in Fig.~\ref{fig:context}. With respect to increasing the
number of lower-cost, higher-performing fusion concepts, an important outcome is that BETHE projects have significantly
enlarged the space of fusion-commercialization teams and concepts
of interest to private investors. For example, 5 BETHE projects or their spinouts--Realta Fusion, Zap Energy, Commonwealth Fusion Systems (CFS), Thea Energy, and Type One Energy--have all raised significant private investments and are now awardees of the DOE Milestone-Based Fusion Development Program~\cite{FOA_milestone}, which assessed both S\&T and commercialization viability as part of the merit review.

As of late 2024, 9 of the 11 Category-A and B projects are still actively 
advancing their concepts or components, respectively. Five 
university or national-laboratory projects have spun out into privately funded companies: University of Wisconsin$\rightarrow$Realta Fusion, University of Maryland Baltimore County (UMBC)$\rightarrow$Terra Fusion, NK Labs$\rightarrow$Acceleron Fusion, PPPL$\rightarrow$Princeton Stellarators (now Thea Energy), and NRL$\rightarrow$LaserFusionX. (Type One Energy had already spun
out of the University of Wisconsin right before the launch of the BETHE program.)
While all the BETHE projects were responsive to the required techno-economic analysis (TEA) in their original BETHE applications,
e.g., providing analysis to show that their scientific-breakeven
experiments could cost $\lesssim$\$100M (Category~A) or that their
OCC could be $<$\$2B and $<$\$5/W (Category~B), the costs and TEA uncertainties
remain large at this stage of fusion development. 
For details on S\&T advances of the
various projects enabled by BETHE, we refer
the reader to the citations in Table~3.

Category-A (concept development) had the objective of enabling at least two or three lower-cost concepts to reach $T_e,T_i \ge 1$~keV for the first time (for those concepts) and at least one concept to achieve fusion triple product $nT\tau_E \ge 10^{18}$~keV$\cdot$s/m$^3$ for the first time (for that concept). Zap Energy achieved $T_e,T_i\ge 1$~keV~\cite{levitt24,goyon24}, and the Wisconsin High-temperature-superconducting Axisymmetric Mirror (WHAM) experiment~\cite{endrizzi23}, designed and constructed under BETHE, recently
initiated operations with the prospect of achieving $T_e,T_i\ge 1$~keV\@. The other 
Category-A projects all made progress advancing beyond their entry milestones though 
have not yet achieved the next milestone (see Table~\ref{table2} for milestone descriptions).

Category-B (component technology development) had the objective
of lowering the cost of more mature concepts via the development
of new component technologies. CFS~\cite{sanabria24}, Type One~\cite{riva23}, and PPPL$\rightarrow$Thea
all developed innovative magnet technologies to enable potentially
lower-cost and faster development paths toward their eventual
grid-ready demonstrations. The University of Rochester and NRL$\rightarrow$LaserFusionX advanced 
broad-bandwidth solid-state and excimer laser technologies,
respectively, and explored high-gain inertial-fusion target designs that can eventually take advantage of these more advanced laser technologies.

Category-C (capability teams) had the objective of bringing
expertise and capabilities (e.g., diagnostics, modeling codes, etc.\@) of teams from national laboratories
and universities to help Category-A teams accelerate their R\&D 
progress. In addition, the collaborations between the capability teams and
the concept development teams (especially the private companies)
were intended to establish better connectivity between the public
and private sectors in fusion energy development and to build
the foundations for future, expanded fusion public-private partnerships. Capability teams would provide objective,
third-party measurements and assessments of physics performance
that are of value for investors conducting due diligence.
The BETHE capability teams all indeed established multiple collaborations
with concept development teams, with tangible collaborative R\&D outcomes evident in the citations in Table~3. 

\section*{GAMOW}


\subsection*{GAMOW Program Objective}

GAMOW \cite{gamow} was informed in part by responses to a Request for Information
(RFI) released in May, 2019 \cite{rfi}, as well as the program-development workshop mentioned earlier.
The RFI stated:
\begin{displayquote}
{\em While it is impossible to predict precisely what is needed for fusion to be commercially viable over the next few decades, fusion’s market entry may require that both the nameplate generation capacity and total construction cost be well below the assumed 1-GWe and $>$\$5B (2019 dollars) scales described in prior fusion-power-plant studies. As discussed further below, this RFI focuses specifically on the enabling technologies for potential fusion power plants at reduced nameplate capacity and cost. ARPA-E is particularly interested in transformational R\&D opportunities that are not already being pursued by or included in the roadmaps of ongoing DOE fusion programs.}
\end{displayquote}

The GAMOW FOA~\cite{FOA_GAMOW}, announced on Feb.~13, 2020
as a joint program with SC-FES, stated:
\begin{displayquote}
{\em The GAMOW program will support innovative R\&D in a range of fusion enabling technologies toward meeting one or more of the following high-level program objectives:
substantial progress toward demonstrating technical feasibility and/or increases in performance in the technical categories of interest (Section I.D) compared to present state-of-the-art; enabling significant device simplification or elimination of entire subsystems, in one or more commercially motivated fusion energy concepts; reduction in fusion energy system cost (capital and/or operations/maintenance) and/or development time/cost, including those of critical materials and component testing/qualification facilities; improvements in the RAMI,\footnote{Reliability, availability maintainability, inspectability.} safety, and/or environmental sustainability of fusion energy systems.}
\end{displayquote}

\begin{figure}[h]
\centering
\includegraphics[width=0.9\textwidth]{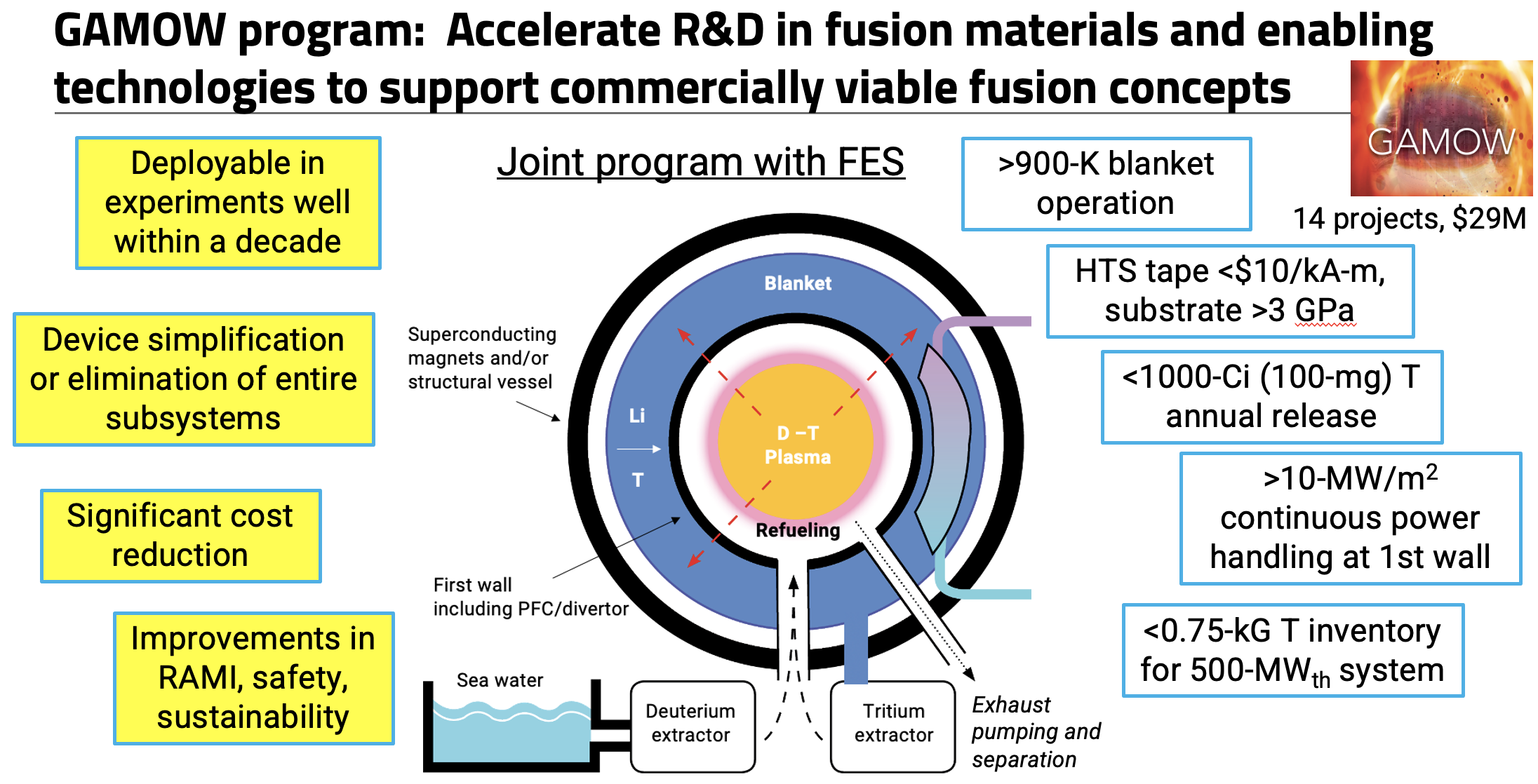}
\caption{Graphic shown at the 2022 ARPA-E Fusion Annual Meeting (April 26--27, 2022 in
San Francisco, CA) summarizing the high-level  objectives and quantitative targets of the GAMOW program.}\label{fig:GAMOW}
\end{figure}

Figure~\ref{fig:GAMOW} summarizes GAMOW's program objectives, technical areas, and performance targets, which are described further below.
The breadth of topics in GAMOW was a reflection of underinvestment
in all these areas relative to ARPA-E's overall objective of 
catalyzing a new trajectory toward timely commercial fusion energy.

\subsubsection*{GAMOW Technical Categories and Performance Targets}

GAMOW applicants were requested to select from one of two research categories, which each had sub-categories listed below. Due to the
range of fusion concepts being pursued and that specific designs would need to make tradeoffs across the different sub-categories, it was not possible to state absolute performance targets across all sub-categories. Nevertheless, minimum quantitative performance targets,
also listed below, were provided for many of the technical sub-categories that would generally improve the techno-economic attractiveness for all D-T fusion concepts. The categories,
sub-categories, and quantitative performance targets were as follows:
\begin{enumerate}
    \item Category A: Fusion Energy Subsystems
    \begin{itemize}
        \item {\em Integrated first-wall and blanket technology}: blanket operating temperature $>900$~K, tritium-breeding ratio $>1.05$
        \item {\em Plasma-facing components and divertor}: continuous power-handling $>10$~MW/m$^2$ in the divertor and $>5$--10~MW/m$^2$ neutron wall load at the first wall
        \item {\em Tritium fuel cycle}: enable tritium site inventory $<0.75$~kg for 500-MWth fusion energy system, environmental annual release limit $\le 1000$~Ci (100~mg T), permeation materials/membranes for tritium extraction $>10^{-7}$~mol m$^{-1}$ s$^{-1}$ Pa$^{-1/2}$, 3--5 times reduction in footprint compared to ITER
        \item {\em High-temperature superconducting (HTS) magnets for fusion}: $>20$~T at $>20$~K, ability to maintain performance up to neutron fluences $>10^{19}$~n/m$^2$, conductor costs $<\$10$/kA-m, yield strength of HTS-magnet substrates $>3$~GPa
        \item {\em High-efficiency electrical-driver systems capable of high-duty-cycle operations}: representative targets of 30~kV, 1~MA, up to 100-$\mu$s flat-top, and 1--10~Hz        
    \end{itemize}
    \item Category B: Cross-Cutting Research
    \begin{itemize}
        \item {\em Novel fusion materials}: Class~C or lower Low-Level Waste at end of life, testing/qualification in prototypic
        neutron environment with up to $10^{18}$~neutrons/m$^2$/s up
        to 150~dpa (displacements per atom)
        \item {\em Advanced and additive manufacturing (AM)}: demonstration
        of AM capabilities for fusion-relevant materials with similar or better properties than those produced by traditional techniques.
    \end{itemize}
\end{enumerate}

\subsubsection*{GAMOW Awardees and Outcomes}

There was a strong response to the GAMOW FOA, resulting
eventually in 14 awardees as summarized in Table~\ref{table:GAMOW}
(see \cite{gamow} for brief project descriptions).
Citations to published papers for each project are provided
in column 2 of Table~\ref{table:GAMOW}. Given that some GAMOW projects have only just recently concluded and others
are still ongoing, we expect to see many more peer-reviewed publications from GAMOW projects in the next few years.

\begin{table}[!ht]
\caption{Summary of GAMOW awardees with citations to published S\&T results.}\label{table:GAMOW}%
\begin{tabularx}{\textwidth}{>{\hsize=.19\hsize}X>{\hsize=.61\hsize}X>{\hsize=.12\hsize}X>{\hsize=.08\hsize\centering\arraybackslash}X}
\toprule
Lead institution & Project title & & Funding\footnotemark[1] \\
\midrule
Category A & & Subsystem & \\
\midrule
Univ.\ California, San Diego & Renewable Low-Z Wall for Fusion Reactors with Built-In Tritium Recovery \cite{martinez23, martinez24} & First wall, divertor & \$1.75M \\ 
SRNL & Process Intensification Scale-Up of Direct LiT Electrolysis & Fuel cycle & \$1.50M\\
Colorado School of Mines & Interfacial-Engineered Membranes for Efficient Tritium Extraction \cite{li23a,job23,li23b,job24,lundin24,li24a,li24b,li25,burst25} & Fuel cycle & \$1.40M \\
SRNL & EM-Enhanced HyPOR Loop for Fast Fusion Fuel Cycles \cite{bliznyuk23,guin24} & Fuel cycle & \$2.30M\\
Univ.\ Houston & Advanced HTS Conductors Customized for Fusion \cite{castaneda23,xue24,paidpilli23,pullanikkat24,sandra25} & Magnets & \$1.50M\\ 
Princeton Fusion Systems & Wide-Bandgap Semiconductor Amplifiers for Plasma Heating and Control~\cite{sen24a,sen24b,vinoth24} & Electrical drivers & \$1.07M\\ 
Bridge12 & High Efficiency, Megawatt-Class Gyrotrons for Instability Control of Burning-Plasma Machines \cite{seltzman23,ridzon23,ridzon24} & Electrical drivers & \$2.75M\\
\midrule
Category B & & Crosscutting area & \\
\midrule
ORNL & Fusion Energy Reactor Models Integrator (FERMI) \cite{bae22,badalassi23,bae23a,bae23b,kropaczek23,borowiec24,bae24} & Integrated modeling & \$3.10M \\ 
UCLA\footnotemark[2] & AMPERE--Advanced Materials for Plasma-Exposed Robust Electrodes \cite{ottaviano23,sabiston24} & Materials & \$1.38M\\ 
Phoenix & Application of Plasma-Window Technology to Enable an Ultra-High-Flux DT Neutron Source \cite{blatz23a,blatz23b} & Materials & \$2.50M \\
ORNL & Advanced Castable Nanostructured Alloys for First-Wall/Blanket Applications & Materials & \$3.30M \\
Stony Brook Univ.\ & ENHANCED Shield: A Critical Materials Technology Enabling Compact Superconducting Tokamaks \cite{snead22,bhardwaj24,dunkin25} & Materials & \$2.55M\\ 
PNNL & Microstructure Optimization and Novel Processing Development of ODS Steels for Fusion Environments \cite{zhang22,wang24} & Materials \& AM & \$2.30M\\
ORNL & Plasma-Facing Component Innovations by Advanced Manufacturing and Design \cite{graning23,robin23,robin24} & Materials  \& AM & \$2.25M\\
\botrule
\end{tabularx}
\footnotetext[1]{SC-FES provided approximately half the funding for GAMOW.}
\footnotetext[2]{The prime recipient is now Oregon State University due to the principal investigator moving there.}
\end{table}

GAMOW projects have made R\&D progress on a range of
fusion materials and enabling technologies, as well as bringing
greater focus on near-term demonstrations of their materials or enabling technologies on integrated fusion experiments. 
Both Category-A and B projects made significant progress toward meeting the major milestones of their respective projects, while improving the prospects for eventually
achieving many of the
program objectives and the performance targets quoted above. In Category A, one example is SRNL's success in creating radiation-resistant, deuterated oils of interest for use in tritium-compatible vacuum pumps as well
as a range of adjacent applications ranging from elastomer seals, lubricants, and IFE foam capsules.
Another is UCSD's discovery and testing of two different composites that show promise as a renewable, slurry-based, first-wall material. In Category B, ORNL's FERMI project identified opportunities to
optimize CFS' liquid-immersion-blanket (LIB) concept and also proposed an alternative blanket concept
called Nested Pebble Bed Blanket (NesPeB); the latter concept was recently selected for follow-on development under the ARPA-E Vision OPEN 2024 program~\cite{open_2021}. Another example is ORNL's 
successful demonstration of the manufacturability of castable nanostructure alloys (CNA) through the production of a 5-ton heat with uniform fine-grain microstructure, a high-density of nanoprecipitates,
and promising initial test results spanning nearly all the pertinent physical properties for
a fusion structural material (publication forthcoming).

For many GAMOW projects, even if the R\&D has progressed to the stage
of readiness for testing or demonstration on an integrated fusion experiment,
there has been a lack of opportunity. Publicly funded experiments
have not prioritized such tests (e.g., due to insufficient
programmatic interest and/or risk tolerance), and privately funded
experiments are still mostly operating at lower levels of fusion performance unsuitable for meaningful
tests.
Certain supply-chain challenges have also been uncovered in
GAMOW, e.g., lack of domestic availability of ton-sized heats of low-activation steel alloys of
the requisite purity and/or compositional tolerances.

Meanwhile, there is rapidly growing private-investor interest in companies pursuing the development of
fusion component technologies, as evidenced by successful private
fundraising by Kyoto
Fusioneering~\cite{KF_2024} and Marathon Fusion~\cite{marathon_arpa-e}) (both of which received ARPA-E funding more recently)
to pursue the development of fuel-cycle, blanket, and/or plasma-heating technologies. Many of the technologies funded through GAMOW may be well-positioned for further commercialization-relevant development through public and/or private funding sources.

\section*{Other Fusion Project Cohorts}

Additional fusion projects were awarded between 2018--2022 under ARPA-E programs other than BETHE and GAMOW\@. These other projects
formed ``cohorts'' with goals consistent with the overall objectives
of the BETHE and/or GAMOW programs. Further details regarding these project cohorts
are summarized in Table~\ref{table:cohorts} and provided in the subsections below.

\begin{table}[!ht]
\caption{Summary of BETHE-GAMOW-era project cohorts awarded under
ARPA-E programs OPEN 2018, Fusion Diagnostic Teams, and OPEN 2021.}\label{table:cohorts}%
\begin{tabularx}{\textwidth}{>{\hsize=.19\hsize}X>{\hsize=.61\hsize}X>{\hsize=.12\hsize}X>{\hsize=.08\hsize\centering\arraybackslash}X}
\toprule
Lead institution & Project title & Concept or Topic & Funding \\
\midrule
OPEN 2018 & & & \\
\midrule
Zap Energy & Electrode Technology Development for the Sheared-Flow Z-Pinch Fusion Reactor \cite{meier21,levitt23,banasek23,shumlak23,levitt24,datta24,goyon24,crews24} & Stabilized Z Pinch & \$6.77M \\ 
CTFusion & HIT-TD: Plasma Driver Technology Demonstration for Economical Fusion Power Plants \cite{sutherland21,morgan22,hossack22} & Spheromak & \$3.42M\\
Princeton Fusion Systems & Novel RF Plasma Heating for Low-Radioactivity Compact Fusion Devices \cite{swanson21,glasser22a,ahsan22,swanson22,glasser22b,vinoth22,dogariu22,galea22,palmerduca22,thomas23,galea23,ahsan23,paluszek23} & FRC & \$1.30M\\ 
\midrule
Fusion Diagnostic Teams & &  & \\
\midrule
Caltech & X-Ray Imaging and Assessment of Non-Perturbing Magnetic Diagnostics for Intermediate-Density Fusion Experiments~\cite{zhou23,zhang23} & High Density & \$400k \\
LLNL & Absolute Neutron Rate Measurements and Non-thermal/Thermonuclear Fusion Differentation~\cite{mitrani21,levitt24,youmans24,conti24,ryan25} & High Density & \$1.33M\\ 
LANL & Portable Soft X-Ray Diagnostics for Transformative Fusion-Energy Concepts~\cite{klemmer23,levitt24} & High Density & \$630k\\
ORNL & A Portable Diagnostic Package for Spectroscopic Measurement of Key Plasma Parameters in Transformative Fusion Energy Devices~\cite{kafle21,kafle22,he22} & Low Density & \$1.60M\\ 
University of Rochester & LLE Diagnostic Resource Team for the Advancement of Innovative Fusion Concepts & High Density & \$1.00M\\
Univ.\ of California, Davis & Electron Density Profile Measurements using Ultrashort-Pulse Reflectometer~\cite{domier21} & Low Density & \$533k\\
LLNL & A Portable Optical Thomson Scattering System \cite{banasek23,levitt24,goyon24} & High Density & \$2.00M\\ 
PPPL & A Portable Energy Diagnostic for Transformative ARPA-E Fusion Energy R\&D & Low Density & \$500k\\
\midrule
OPEN 2021 & & & \\
\midrule
LANL\footnotemark[1] & Advanced Manufacturing of High-Entropy Alloys as Cost-effective Plasma Facing Components for Fusion Power
Generation~\cite{hatler25} & Fusion Materials & \$3.11M \\
MIT & Liquid Immersion Blanket: Robust Accountancy~\cite{ferry22,meschini23,delaporte25} & Blanket & \$3.06M\\
Princeton Univ. & Economical Proton-Boron11 Fusion~\cite{ochs22,kolmes22,mlodik23,ochs23a,munirov23,kolmes23,ochs23b,rax23b,ochs24a,kolmes24a,kolmes24b,ochs24b,kolmes24c,rubin24,ochs24c,kolmes24d,gueroult24} & p-$^{11}$B & \$2.47M\\
Polymath Research & Longer Wavelength Lasers for Inertial Fusion Energy with Laser-Plasma Instability Control: Machine Learning
Optimum Spike Trains of Uneven Duration and Delay (STUD Pulses) & Laser Fusion & \$2.01M\\
\botrule
\end{tabularx}
\footnotetext[1]{The prime recipient is now PNNL due to the principal investigator moving there.}
\end{table}

\subsection*{OPEN 2018 and OPEN 2021}

Every three years, ARPA-E has announced an ``open'' FOA that complements its focused technology programs (such as BETHE and GAMOW), inviting applications that broadly address
ARPA-E's statutory mission in new and emerging areas across the complete spectrum
of energy applications. OPEN 2018~\cite{FOA_OPEN2018} and OPEN 2021~\cite{FOA_OPEN2021} were the fourth and fifth OPEN FOAs, respectively, in the history of ARPA-E. Three fusion projects were awarded under OPEN 2018 and four under OPEN 2021, as summarized in Table~\ref{table:cohorts}.

\subsubsection*{Awardees and Outcomes of the OPEN 2018 Fusion Cohort}

The fusion projects under OPEN 2018 (see Table~\ref{table:cohorts}) represented an expansion of ARPA-E support into fusion 
concept development beyond the MIF concepts of ALPHA\@. While Zap Energy was a spin-out and
follow-on to the University of Washington ALPHA project on the sheared-flow stabilized (SFS) Z pinch,
CTFusion and Princeton Fusion Systems added the sustained spheromak and field-reversed
configuration (FRC), respectively, to ARPA-E's portfolio of fusion approaches.
The objectives and metrics for the fusion cohort under OPEN 2018 largely adhered to those of Category A of the BETHE program, as described earlier in this paper. For detailed technical results of
these projects, we refer the reader to the references given in Table~\ref{table:cohorts}.

Zap Energy continues to make progress in advancing the fusion performance of the SFS Z pinch, including
recently becoming one of just a handful of fusion concepts to demonstrate electron temperature
exceeding 1~keV~\cite{levitt24}, as well as one of eight awardees of the DOE Milestone-Based Fusion Development Program~\cite{FOA_milestone}. Cumulatively, Zap Energy has raised more than \$330 million of private funding since spinning out of the University of Washington in 2017.

CTFusion made good progress on its project milestones, including theory/modeling
of the sustained spheromak, collecting new data on the existing experimental device HIT-SI3 
(Helicity Injected Torus - Steady Inductive 3) at the
University of Washington to inform design of the next device, and on the design, construction, and initial operations of an upgrade to HIT-SI3 called HIT-SIU\@.
Unfortunately, pandemic-induced delays permitted only 4--5 months of HIT-SIU experimental operations before project funding ended.
The viability and scalability of steady-state, inductive, helicity injection for spheromak sustainment remains an open question.

Princeton Fusion Systems also made progress on its project milestones, including
diagnostic measurements and upgrades to the FRC experiment located at PPPL\@. 
Pandemic-induced delays precluded the project from meeting its key milestones of 
demonstrating elevated electron and ion heating at optimized frequencies of its odd-parity,
rotating-magnetic-field (RMF) approach to FRC heating and sustainment. The viability
and scalability of odd-parity RMF for FRC heating and sustainment remains an open question.

\subsubsection*{Awardees and Outcomes of the OPEN 2021 Fusion Cohort}

The ongoing fusion projects under OPEN 2021 (see Table~\ref{table:cohorts}) further expanded ARPA-E's 
fusion portfolio into new areas of relevance to the program objectives of both BETHE and GAMOW\@.
Because these projects are still ongoing, we anticipate many more peer-reviewed publications over the next few years.
The Princeton project has been prolific in exploring fundamental
and novel plasma processes (e.g., rotational flows,
control of ion velocity distributions, wave-particle interactions, and
suppression of bremsstrahlung, etc.)\ that may be collectively required to establish plasma
equilibria capable of sustaining fusion energy gain in a proton-boron
fusion plasma. The objective of this ongoing work is to identify such
a theoretical equilibrium to justify and motivate subsequent experimental
testing and validation. The Polymath project is developing and training models based on large
laser-plasma experimental datasets to enable future control
and mitigation
of laser-plasma instabilities (LPI) for IFE, via a technique
known as STUD (spike trains of unaaaaeven duration and delay) pulses.
The LANL project has focused on alloy design, development,
testing, and additive manufacturing of W-based high-entropy alloys 
for plasma-facing components that could ultimately enable higher 
radiation resistance, as well as lower capital and operational costs
of future fusion power plants.
The MIT project has designed and is building the LIBRA (Liquid Immersion
Blanket: Robust Accountancy) experiment, which is a sub-scale, integrated 
experiment to address priority research questions for molten
FLiBe blankets and tritium breeding.

\subsection*{Fusion Diagnostic Teams}

The Fusion Diagnostic Teams topic~\cite{FOA_TINA} was announced
in Feb.~2019 to support the validation of achieved experimental
performance of ALPHA and OPEN~2018 fusion projects, as well as to
to support the anticipated
BETHE program (announced in Nov.~2019). The FOA (Appendix~D)~\cite{FOA_TINA} stated:
\begin{displayquote}
{\em The purpose of this announcement is to solicit submissions from experienced fusion researchers who have designed, implemented, and operated diagnostics for the characterization of high-temperature fusion plasmas. The immediate objectives of this Targeted Topic are:
\begin{itemize}
\item[a.] Realizing plasma diagnostic systems, especially more costly or complex systems, that can be transported to and shared among different fusion experiments;
\item[b.] Enabling high-quality diagnostic measurements of plasma and fusion parameters on fusion-concept-exploration experiments supported by ARPA-E;
\item[c.] Leveraging the plasma diagnostic expertise of the entire fusion R\&D community to advance potentially transformative fusion-concept research; and
\item[d.] Developing teams and experience to support a potential future expansion of diagnostic resource teams for fusion and private-public partnerships in general.
\end{itemize}

ARPA-E supports the exploration and development of potentially transformative fusion-energy concepts, with key aims of significantly lowering development costs and accelerating the timeline to commercial fusion energy. Most of the ARPA-E-supported fusion experiments would benefit from definitive diagnostic measurements to firmly establish the level of performance that has been achieved and/or to clearly identify issues if measured performance is worse than expected. In particular, several of these concepts are showing evidence of thermonuclear neutron production that could be consistent with ion temperatures above 1~keV (approximately 10~million degrees K). Measurements of key plasma parameters in these experiments are needed to confirm the observed neutron yields and determine whether they are consistent with
thermonuclear plasma conditions. Such measurements would improve our understanding of these concepts and help identify and correct any deficiencies, if necessary. All of the concepts, including those not yet at keV-level temperatures, would benefit from multi-point, spatially and temporally resolved measurements of plasma parameters and their spatial profiles to experimentally infer particle, energy, and magnetic-field transport.

Many of the needed diagnostics require significant expertise and/or expensive hardware that are beyond the typical resources available to an earlier-stage fusion project. The diagnostics (and the resources to operate them and analyze the data) can cost as much as or more than the fusion experiment itself, and typically these diagnostics are available only at national-scale fusion facilities. A primary aim of this Targeted Topic is to stretch limited resources such that multiple promising, earlier-stage fusion experiments can benefit from advanced measurements to validate their performance, uncover problems, and guide research priorities.}
\end{displayquote}

The FOA sought transportable diagnostics with the
technical parameters given in Table~\ref{table:diagnostics} for
magnetically confined and pulsed, intermediate-density ARPA-E-funded
fusion concepts.

\begin{table}[!ht]
\caption{Diagnostic parameters of interest in the Fusion Diagnostic Teams
FOA~\cite{FOA_TINA}.}\label{table:diagnostics}%
\begin{tabularx}{\textwidth}{>{\hsize=.4\hsize}X>{\hsize=.3\hsize\centering\arraybackslash}X>{\hsize=.3\hsize\centering\arraybackslash}X}
\toprule
Parameter & Magnetically confined & Pulsed, intermediate density \\
\midrule
Ion and electron density (cm$^{-3}$) & $10^{13}$--$10^{14}$ & $10^{16}$--$10^{21}$ \\
Electron temperature (eV) & 10--2000 & 100-3000\\
Ion temperature (eV) & 10--2000 & 100--10000\\
Magnetic field (T) & 0.1--3 & 1--1000\\
Neutron yields & N/A\footnotemark[1] & $10^6$--$10^{12}$ (per pulse) \\
Neutron energy (MeV) & N/A\footnotemark[1] & 2.3--2.8 w/few-keV resolution\\
Neutron duration (ns) & N/A\footnotemark[1] & 10--10000\\
Time resolution & $<100$~$\mu$s & 1--1000~ns\\
Spatial resolution & $<1$~cm & $<1$~mm\\
\botrule
\end{tabularx}
\footnotetext[1]{Magnetically confined concepts supported by ARPA-E
were not expected to be generating meaningful neutrons yields during the period of
performance of the Fusion Diagnostic Teams program.}
\end{table}

\begin{figure}[h]
\centering
\includegraphics[width=0.9\textwidth]{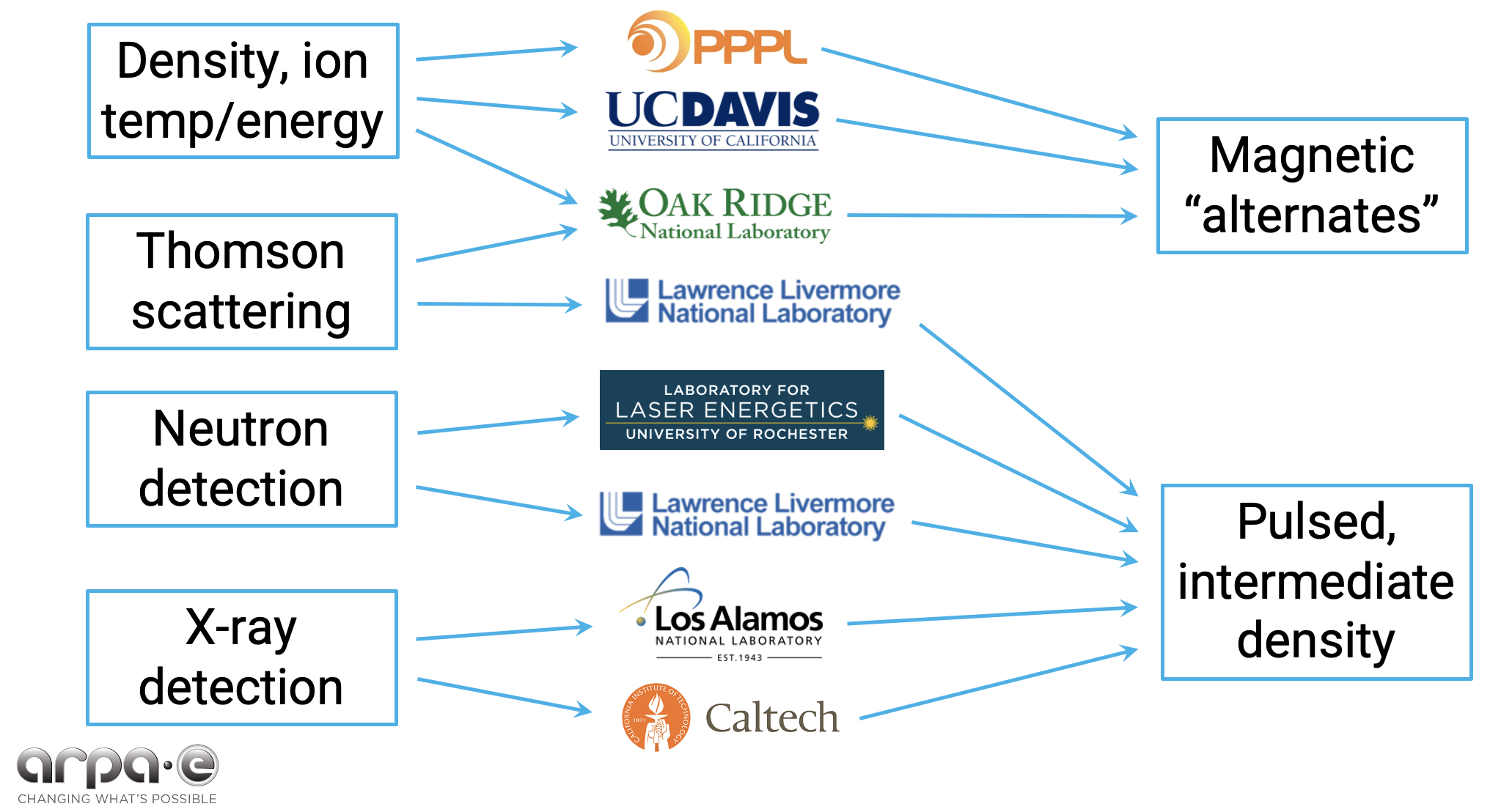}
\caption{Summary graphic shown at the (virtual) FUSION Diagnostics First Annual Review Meeting (March 5, 2021).}\label{fig:diagnostics}
\end{figure}

\subsubsection*{Awardees and Outcomes of the Fusion Diagnostic Teams Cohort}

Selected projects are listed in Table~\ref{table:cohorts} and summarized
in Fig.~\ref{fig:diagnostics}. All of the projects were successful
in developing new, transportable diagnostics based on established
diagnostic approaches. Most of the projects were successful in
deploying their diagnostic(s) to at least one ARPA-E-supported
fusion experiment. Of particular note are the LLNL and LANL teams
deploying their diagnostics to Zap Energy's experiments that
were supported under OPEN~2018 and BETHE, leading
to successful measurements and confirmation of the achievement
of plasma parameters of significance for fusion development, e.g.,
\cite{levitt24,goyon24}. Several teams have received subsequent
SC-FES INFUSE (Innovation Network for Fusion Energy)~\cite{INFUSE} awards to support the design and/or deployment of their
diagnostics for other fusion experiments. Many of these teams and diagnostics continue to be a resource for both privately and publicly supported fusion experiments.

\section*{BETHE-GAMOW-Era T2M Activities}

All ARPA-E focused programs are accompanied by a technology-to-market (T2M) effort led by a T2M Advisor, working closely with the Program Director, to support the overall program
and its individual project teams in their commercialization pathways.

Earlier T2M efforts of the ALPHA program~\cite{nehl19} examined
the fusion patent landscape, supported the development of a
fusion costing model and application of the model to fusion concepts
in the ALPHA program, commissioned a JASON study to examine
and make recommendations regarding the continuing prospects of MIF,
and strengthened outreach to the scientific community by supporting
ALPHA teams in organizing a miniconference in the 2018 Annual
Meeting of the American Physical Society (APS), Division of Plasma Physics.

The BETHE-GAMOW-era T2M efforts built on those of the ALPHA program,
focusing on investor engagements, continued support of fusion costing modeling,
identifying potential early markets for fusion energy and
their cost targets, drawing attention
to topics of importance for future fusion demonstration and deployment (e.g.,
social license and waste disposition/management), and outreach activities
especially presenting to and engaging with students at universities.

\subsection*{Investor Engagements}

Investor engagement became an initial fusion-T2M priority during the BETHE-GAMOW era 
because of the stated BETHE program goals of attracting further private investments
and building the foundations for sustainable public-private partnerships to develop fusion
energy.
When the BETHE program was being developed (c.~2019), private-investor interest in fusion was limited (e.g., many investors did not respond to our emails) but
growing (e.g., Commonwealth Fusion Systems announced its \$115M Series~A in June 2019~\cite{CFS_series_A}).
Our investor engagements focused on (1)~better understanding the
investment theses of a wide range of investors (i.e., those who already invested in fusion, those who considered but declined,
and those interested but had not yet invested); (2)~better understanding
of what types of government support could effectively catalyze
greater private fusion investments; and (3)~educating
potential fusion investors on key S\&T challenges of fusion
development and the techno-economic tradeoffs they might be making by investing
in certain fusion companies versus others. During the BETHE-GAMOW era, we spoke to
$\sim$100 investors (representing a wide range
of asset managers including venture capital, university endowments, sovereign wealth funds, state retirement funds, family offices, investment banks, and high-net-worth individuals or their representatives), invited some of them to speak at ARPA-E fusion workshops and meetings, and 
published a paper~\cite{wurzel22} (with investors in mind as a primary
audience) describing the importance of the Lawson criteria
as an objective measure of scientific progress toward fusion energy breakeven.

\subsection*{Fusion Techno-Economics}

True to one of ARPA-E's mantras {\em if it works, will it matter?},
fusion techno-economics was another focus area of the BETHE-GAMOW-era fusion-T2M efforts. We wanted to increase clarity and raise
attention on fusion costs and
required techno-economics for it to be economically viable in the
competitive energy markets of the future. Furthermore, 
knowledge of the required techno-economics is critical to guide ARPA-E's fusion R\&D priorities. Thus, during the BETHE-GAMOW era, our fusion-T2M efforts included techno-economic activities on improving the cost modeling for fusion plants, as well as on identifying levelized-cost-of-electricity (LCOE) targets for fusion
to gain entry into various energy markets of the future.

In 2019--2020, ARPA-E supported Woodruff Scientific and collaborators
to update the earlier costing study
performed during the ALPHA program. The updated study~\cite{WSI_updated_study_2020}
revisited the overnight construction cost (OCC) for four fusion concepts
supported under the ALPHA program and added LCOE estimates. Since
then, ARPA-E supported Woodruff Scientific to create
open-source python libraries of the costing models~\cite{Woodruff_Scientific}. Although the estimated OCC and LCOE
values have large uncertainties at this point due to fusion's relatively
low technological readiness level (TRL), the costing models
are extremely useful for identifying significant cost drivers and
for assessing the cost impacts of different S\&T choices in the design of a fusion plant.
Woodruff Scientific has applied the costing model to many privately funded fusion concepts including those outside of ARPA-E's fusion
programs.

Equally importantly, energy markets and competing
technologies will determine the required costs for fusion to
gain market entry. To support ARPA-E's fusion performers
(and fusion developers writ large)
in market analysis, we published a paper~\cite{handley21} that explored a range of potential
early markets for fusion energy and identified approximate cost targets for fusion to gain market entry (including both electricity
and non-electricity energy applications and accounting for
geographic differences). Key findings were that fusion developers should consider focusing initially on high-priced electricity markets and including integrated thermal storage, depending on techno-economic factors, in order to maximize revenue and compete in markets with high penetration of renewables. Process heat and hydrogen production may be tough early markets for fusion, but may open up to fusion as markets evolve and if fusion’s LCOE falls below 50~\$/MWh$_e$. At the time of the paper's publication (2020), the demand
for carbon-free electricity to power the rapid growth in data
centers for artificial intelligence had not yet arrived, but
this may be a potential early market for fusion energy as well.
Finally, the paper~\cite{handley21} discussed the potential ways for a fusion plant to increase revenue through cogeneration (e.g., desalination, direct air capture or district heating) and to reduce capital costs (e.g., by minimizing construction times and interest or by retrofitting coal plants).

\subsection*{Social License}

Public acceptance is critical for the deployment of any energy
technology, and we were fortunate to be engaged with experts who
helped us communicate the importance to ARPA-E fusion performers
(and the broader fusion R\&D community) of earning a ``social license'' for fusion. Fusion may encounter headwinds with respect to public acceptance
because of, among other things, the need to process/contain unprecendented amounts of tritium and
the generation of large volumes of activated structural materials
requiring short-term storage and viable recycling pathways.

Jane Hotchkiss, formerly of Pegasus Fusion Strategies and now with Energy for the Common Good, emphasized
the importance of developing partnerships with nongovernmental,
public-advocacy, and philanthropic groups that will help fusion
earn long-term public acceptance. We invited her to speak on this topic
at the ARPA-E fusion program-development workshop in August 2019 and to
lead a panel discussion
on this topic at the 2020 BETHE kickoff meeting.
Throughout the BETHE-GAMOW era, the ARPA-E fusion team engaged with organizations such as the
Natural Resources Defense Council, Stellar Energy Foundation, Electric Power Research Institute, Clearpath, Clean Air Task Force, Information Technology and Innovation Foundation, Schmidt Futures, and Emerson Collective.

Dr.~Seth Hoedl of the Post Road Foundation emphasized the importance
of achieving a social license for fusion in his response to our 2019 RFI~\cite{rfi}, and we invited him to speak on the topic
at the GAMOW kickoff meeting in January 2021. Shortly thereafter,
we nominated Dr.~Hoedl to present his recommendations
to the larger plasma and fusion R\&D community in an invited
talk~\cite{hoedl22} at the 2021 Annual Meeting of the APS Division of Plasma Physics. Dr.~Hoedl emphasized the importance of learning
from successful public engagement strategies and learning
about international views on fusion that may be significantly different from our own. To further these discussions, we invited Dr.~Hoedl to help us
co-organize a ``Social Acceptance Mini-Workshop'' at the 2022
ARPA-E Fusion Annual Meeting.

The disposition/management of fusion's activated waste and byproduct materials may
impact public acceptance for fusion, as well as fusion's techno-economics. Thus, we also invited Dr.~Laila El-Guebaly (Univ.
of Wisconsin-Madison, retired) to speak
on fusion waste disposition/recycling at the 2021 GAMOW kick-off
meeting in January 2021,
as part of our efforts to raise awareness among
ARPA-E's fusion performers (and the fusion R\&D community writ large) on broader issues of importance for fusion commercialization.

\subsection*{Project-Level T2M Support}

We provided project-level T2M support for every
ARPA-E fusion project. This involved guiding project teams in
developing and updating their T2M strategies that focused both on their longer-term objectives of realizing or enabling commercially viable fusion energy, as well as nearer-term objectives of sustaining their development
and commercialization pathways after the end of their ARPA-E projects.
The longer-term T2M objectives could focus, e.g., on cost modeling
and market analysis for performers' specific fusion concepts
or component technologies. The nearer-term T2M objectives often
focused on next-step funding, which included coaching performers
on preparing their investor pitch decks, introducing
performers to investors and/or other potential partners, and, for a small number
of performers, providing dedicated T2M funds to help teams spin-out
and establish a private company to raise equity investments.
Select project-level T2M notes (beyond what was discussed earlier in this paper) are provided in Appendix~A\@.

\section*{Conclusions}

ARPA-E's BETHE-GAMOW-era fusion 
programs and project cohorts, along with the emergence of the Fusion Industry Association,\footnote{Seeds for the formation of the Fusion Industry Association in 2018~\cite{FIA_launch}
were in part planted through discussions arising from the
ALPHA-era ARPA-E fusion annual meetings.} the
2020 Fusion Energy Sciences Advisory Committee Long-Range Plan {\em Powering the Future}~\cite{FESAC_LRP}, and the 2021 U.S.
National Academies report {\em Bringing Fusion to the U.S. Grid}~\cite{NASEM21}, together
contributed to
a policy shift in U.S. fusion-energy development starting in 2022. The new
emphasis and aspiration is to partner with the private
sector to realize industry-led fusion pilot plants in the 2030s~\cite{hsu23}.

Many BETHE-GAMOW-era fusion projects remain active, and thus the
full outcome and impacts (S\&T and otherwise) may not be fully known for several
years or more. It was beyond the
scope of this paper to provide a detailed review of the
S\&T outcomes of the BETHE-GAMOW-era fusion projects, but
this paper's reference list includes
a record of the published R\&D results to-date (in over 120 peer-reviewed journal articles). Many additional manuscripts are in preparation or under
review with journals. We are aware of many more patents or patent disclosures by BETHE-GAMOW-era project teams than are reflected in iEdison (the interagency online reporting system for federally funded subject inventions) at this time, and we urge all ARPA-E fusion performers to update iEdison with subject inventions arising from ARPA-E funding.
As of May 2025, BETHE-GAMOW-era projects or their spin-outs have cumulatively raised more than \$500M in equity investments to date (not including CFS) since their selection into one or more of these programs or project cohorts. 

Program Director Dr.~Ahmed Diallo joined ARPA-E in 2022 (succeeding Program Director Dr.~Scott Hsu) and inherited management of
most of the active ARPA-E fusion portfolio shortly after joining. In 2024, Dr.~Diallo launched the CHADWICK (Creating Hardened and Durable Fusion First-Wall Incorporating Centralized Knowledge) program~\cite{chadwick}
that seeks to reimagine what is possible in first-wall and
structural materials to better enable the commercial viability of future fusion power plants.

\backmatter




\bmhead{Acknowledgements}

The ARPA-E fusion team during the BETHE-GAMOW era included
Scott Hsu (Program Director, 2018--2022), Ahmed Diallo (Program
Director, 2022--present), Robert Ledoux (Program Director, 2020--2025), Malcolm Handley (T2M Consultant, 2019--2020), Sam Wurzel
(T2M Advisor, 2021--2024), Heather Jackson (T2M Advisor, 2024--present), Colleen Nehl (Booz
Allen Hamilton, lead fusion Tech SETA,\footnote{Science, Engineering, Technology Advisor.} 2014--2022), and Ed Cruz (Booz Allen Hamilton, lead fusion Tech SETA, 2022--present). Patrick McGrath
was the ARPA-E Deputy Director for Technology (2018--2020), Associate Director of Technology (2016--2018), and Program Director for ALPHA (2014--2018). The team members above who are presently federal employees or contractors
could not be listed as co-authors at this time due to a pause on
publications by federal employees and contractors. 

We also thank the many additional people who contributed tirelessly to the execution of the
BETHE-GAMOW-era fusion programs and project cohorts: Aron Newman, Curt Nehrkorn, Pankaj Trivedi, Eric Carlson, Zia Rahman, Mark Pouy, Maruthi Devarakonda, Robert Thompson, and Christina Leggett (Booz Allen Hamilton Tech SETAs), who supported technical aspects of the day-to-day project 
management of the nearly 50 fusion projects of this era; David Hanlin, Raphael Wineburg, and Whitney White (Booz Allen Hamilton Program SETAs), who supported the budgetary and
contractual aspects of the day-to-day project management; Kramer Akli, Josh King, and Sam Barish 
(SC-FES Program Managers) for their partnership in supporting three of the BETHE projects; and Guinevere Shaw and Daniel Clark (former SC-FES Program
Managers), who collaborated closely on the joint ARPA-E/FES GAMOW program. We gratefully acknowledge all the ARPA-E fusion
performers, whose 
dedication, creativity, and passion brought these programs to 
life and advanced the prospects of timely commercial fusion energy.

The program names BETHE and GAMOW are a tribute to the towering scientific contributions in fusion by Hans Bethe~\cite{bethe39} and George Gamow~\cite{gamow28}. The triumvrate of ARPA-E's first three fusion programs ALPHA, BETHE, and GAMOW pays tribute to another famous paper relevant to fusion~\cite{alpha48}.

\begin{appendices}

\section{Select Project T2M Notes}\label{sec:anecdotes}

\subsection*{PPPL (BETHE)}

The PPPL ``Stellarator Simplification using Permanent Magnets'' BETHE project faced a neodymium price increase of over 200\% from 2019 to 2022, rendering their plan to construct a toroidal sector of a neodymium-based, permanent-magnet stellarator far outside of budget. This in part contributed to a reconsideration of their approach and the realization that arrays of planar HTS coils could obviate the need for permanent magnets. The technical benefits and potentially improved economics of this scheme led the team to abandon the use of permanent magnets in favor of arrays of HTS coils and spinning off Thea Energy to pursue the planar-coil stellarator concept. Thea Energy closed a \$20M series A in 2024. Although the original project to build and characterize a toroidal sector of a permanent-magnet assembly was not completed, this was a successful project from both technical and commercialization perspectives.

\subsection*{University of Wisconsin-Madison (BETHE)}

Novel applications of HTS played a key role in the spinout of Realta Fusion from the ``University of Wisconsin-Madison HTS Axisymmetric Magnetic Mirror'' (WHAM) project. Investors grasped the inherent economic advantages of a linear fusion machine and the transformational impact of much larger magnetic fields enabled by HTS\@. Realta Fusion raised a \$9M equity seed round in 2023 and a \$36M Series~A in 2025. The HTS magnets (sourced from Commonwealth Fusion Systems) were successfully installed and operated on WHAM in 2024. WHAM is the highest magnetic-field, magnetic-confinement device ever built thus far, and it is worth noting that such a world-class machine (currently attracting top graduate students and researcher collaborators) was built for less than \$20M in total and \$10M of federal funding.

\subsection*{University of Maryland, Baltimore County (BETHE)}

Limited budgets drove creative design decisions during the development of the Centrifugal Mirror Fusion Experiment at the University of Maryland, Baltimore County. They purchased three decommissioned magnetic-resonance-imaging (MRI) magnets at low cost. One was disassembled to precisely characterize the 
design and magnetic field pattern, and the other two were installed as the primary mirror coils of the experiment. This repurposing of used MRI magnets resulted in lower costs and shorter construction times. The team has spun out the company Terra Fusion.

\subsection*{Collaboration among LLNL, LANL, and Zap Energy}

The LLNL (Fusion Diagnostics topic) and LANL (Fusion Diagnostics topic and BETHE) diagnostic teams, which measured electron temperature, soft-x-ray emission, and neutron yield on Zap Energy's FuZE experiment, bolstered confidence (by investors and the broader fusion research community alike) due to published results from this experiment in top-tier research journals, e.g., \cite{levitt24,goyon24}. These third-party measurements, combined with the strong economic case for a fusion concept without external magnets, resulted in over \$330M of equity investments into Zap Energy. This is a remarkable sum relative
to the cumulative ARPA-E support for the SFS Z-pinch (starting with University of Washington in the ALPHA program~\cite{nehl19} and continuing under OPEN~2018 and BETHE) and the associated work of the LLNL/LANL diagnostic teams of less than \$15M.



\subsection*{NK Labs (BETHE)}

NK Labs’ spin-off Acceleron Fusion captured the imagination of investors with a straightforward pitch explaining new opportunities (both scientific and technological) that could potentially counter the reasons why a muon-catalyzed-fusion power plant appeared impossible in previous decades (i.e., the ``muon sticking problem'' and the high cost of muons from the earlier generation of accelerators). The Acceleron team explained how the high-pressure fuel regime they are exploring (supported under BETHE) might have a lower sticking rate and described potential pathways to lower-cost muon production using modern accelerators. Based on this pitch, Acceleron closed a \$24M Series A funding round in 2024.


\subsection*{Commonwealth Fusion Systems (BETHE)}

Commonwealth Fusion System’s (CFS) fundraising activities were well underway at the time of their BETHE award to develop a fast-ramping HTS central solenoid. Not only was this project successful in demonstrating an HTS central solenoid model coil (underpinning the use of this technology in SPARC), but work on the underlying VIPER cable also found application in the non-planar stellarator coils under development (also initiated under BETHE) at Type One Energy.

\subsection*{Type One Energy (BETHE)}

Type One Energy applied to the BETHE program with the idea
to build non-planar HTS stellarator coils leading to an HTS stellarator power plant. Their BETHE project applied advanced manufacturing to this challenge, resulting in successful technical tests of bent stellarator cables and metal, 3D-printed supports. Beyond these technical accomplishments, the company expanded its management team, raised \$87M of private funding, and gained entry into the DOE’s Milestone-Based Fusion Development Program~\cite{FOA_milestone}.

\subsection*{University of Rochester and NRL (BETHE)}

University of Rochester and NRL’s IFE target and driver development project resulted in two spin-off companies, PolyKrom Fusion and LaserFusionX, respectively. PolyKrom was unable to raise equity funding, partially due to disagreements between the company and the University of Rochester and other issues. LaserFusionX continues to pursue equity funding to advance their direct-drive IFE concept based on excimer lasers.




\end{appendices}


\end{document}